\documentclass[aps,prl,twocolumn,superscriptaddress,showpacs,preprintnumbers]{revtex4}

\usepackage{graphicx}

\begin{document}

\title{Fermi Surface and Band Renormalization in (Sr,K)Fe$_2$As$_2$ Superconductor from Angle-Resolved Photoemission Spectroscopy}
\author{Haiyun Liu$^{1}$,  Wentao Zhang$^{1}$, Lin Zhao$^{1}$, Xiaowen Jia$^{1}$, Jianqiao Meng$^{1}$,
Guodong  Liu$^{1}$, Xiaoli Dong$^{1}$, G. F. Chen$^{2}$, J. L.
Luo$^{2}$,  N. L. Wang$^{2}$, Wei Lu$^{1}$, Guiling Wang$^{3}$, Yong
Zhou$^{3}$, Yong Zhu$^{4}$, Xiaoyang Wang$^{4}$, Zuyan Xu$^{3}$,
Chuangtian Chen$^{4}$, X. J. Zhou$^{1,*}$}

\affiliation{
\\$^{1}$National Laboratory for Superconductivity, Beijing National Laboratory for Condensed
Matter Physics, Institute of Physics, Chinese Academy of Sciences,
Beijing 100190, China
\\$^{2}$Beijing National Laboratory for Condensed Matter Physics, Institute of Physics,
Chinese Academy of Sciences, Beijing 100190, China
\\$^{3}$Key Laboratory for Optics, Beijing National Laboratory for Condensed Matter Physics,
Institute of Physics, Chinese Academy of Sciences, Beijing 100190,
China
\\$^{4}$Technical Institute of Physics and Chemistry, Chinese Academy of Sciences, Beijing 100190, China
}
\date{July 20, 2008}
%
%

\begin{abstract}

High resolution angle-resolved photoemission measurements have been
carried out on (Sr,K)Fe$_2$As$_2$ superconductor (T$_c$=21 K). Three
hole-like Fermi surface sheets are resolved for the first time
around the $\Gamma$ point which is consistent with band structure
calculations. One electron-like Fermi surface and strong Fermi spots
are observed near the M($\pi$,$\pi$) point.  The overall electronic
structure, particularly near the M point, shows significant
deviations from the band structure calculations. The obvious
bandwidth renormalization suggests the importance of electron
correlation in understanding the electronic structure of the
Fe-based compounds.

\end{abstract}


\maketitle

The recent discovery of superconductivity in iron-based ReFeAs(O,F)
(Re represents rare earth elements like La,Ce,Pr,Nd,Sm and etc.)
\cite{Kamihara,XHChen43K,GFChenCe,ZARenNd,ZARenPr,ZARenSm} and
(A,K)Fe$_2$As$_2$ (A represents alkaline earth elements like Ba and
Sr)\cite{RotterSC,Sasmal,GFCSrFeAs,XHCBaFeAs} has attracted great
attention because they represent second class of high temperature
superconductors after the  discovery of first high temperature
superconductivity in cuprates\cite{Bednorz}. Different from the
cuprates where the parent compound is a Mott
insulator\cite{PLeeReview}, the parent compounds of the iron-based
superconductors show a metallic behavior with a spin-density-wave
ground state\cite{DongSDW,PCDai,BFSSDW}. This has raised an
important question on whether one should treat iron-based compounds
with an itinerant electron model\cite{DJSingh,HJZhang} or localized
correlated model\cite{KHaule,CCao,ZPYin,FJMa}.  Direct measurement
of the electronic structure is crucial in addressing this issue, and
particularly the effect of electron correlation in this iron-based
system\cite{KaminskiLOFA,DLFKFe2As2,CLiuKFeAs}.

In this paper, we report first direct measurements of the Fermi
surface and band structure of the (Sr$_{1-x}$K$_x$)Fe$_2$As$_2$
superconductor by angle-resolved photoemission (ARPES) measurements.
We have clearly identified three hole-like Fermi surface sheets near
the $\Gamma$ point of the Brillouin zone, which is consistent with
the band structure calculations. We also observe an electron-like
Fermi surface and strong Fermi spots near the M($\pi$,$\pi$) point.
The overall electronic structure, particularly around the M point,
shows significant difference from the band calculations.  In
addition, we observed an obvious bandwidth renormalization which
suggests the importance of electron correlation in understanding the
electronic structure of the iron-based compounds. These results
provide important information in establishing the basic electronic
structure of the iron-based high temperature superconductors.

\begin{figure*}[floatfix]
\begin{center}
\includegraphics[width=1.6\columnwidth,angle=0]{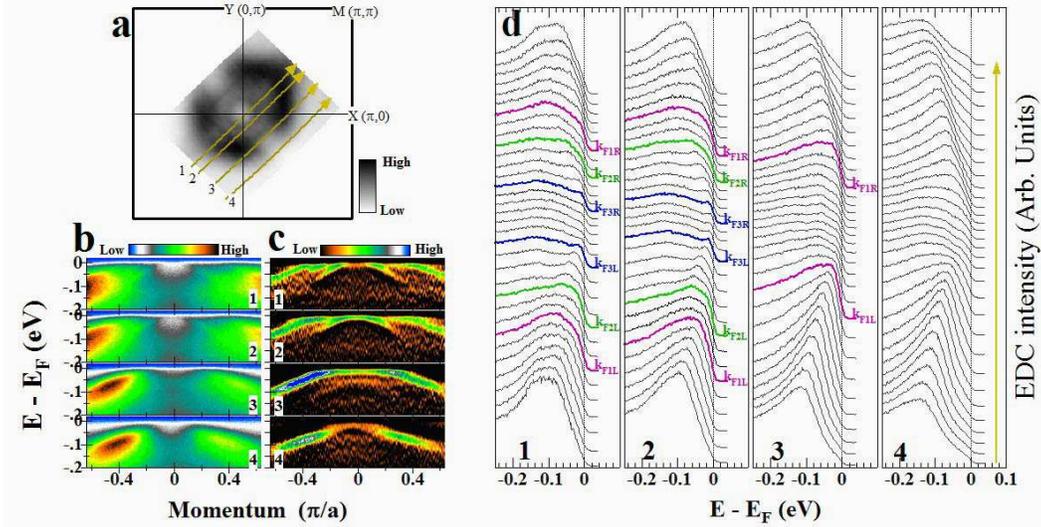}
\end{center}
\caption{Fermi surface, band structure and photoemission spectra of
(Sr,K)Fe$_2$As$_2$ (T$_c$=21 K) near the $\Gamma$ point measured at
45 K. (a). Spectral weight integrated within [-5meV,5meV] energy
window with respect to the Fermi level as a function of k$_x$ and
k$_y$. The black arrow near the bottom-right marks the main electric
field direction on the sample surface from the light source. (b).
Original photoemission images measured along the four typical cuts
as marked in Fig. 1a. (c). Corresponding second derivative images of
Fig. 1b. (d). Photoemission spectra along the four cuts with EDCs at
the Fermi momenta colored and marked. }
\end{figure*}

The angle-resolved photoemission measurements are carried out on our
lab system equipped with Scienta R4000 electron energy analyzer with
wide angle mode (30 degrees)\cite{GDLiu}. We use Helium I resonance
line as the light source which gives a photon energy of h$\upsilon$=
21.218 eV. The light on the sample is partially polarized with the
electric field vector mainly in the plane of the sample surface (as
shown in Fig. 1a). The energy resolution was set at 12.5 meV and the
angular resolution is $\sim$0.3 degree. The Fermi level is
referenced by measuring on the Fermi edge of a clean polycrystalline
gold that is electrically connected to the sample. The
(Sr$_{1-x}$K$_x$)Fe$_2$As$_2$ single crystals were grown using flux
method \cite{GFChenCrystal} and the crystal measured has a
superconducting transition at T$_c$=21 K\cite{Composition}. The
crystal was cleaved {\it in situ} and measured in vacuum with a base
pressure better than 6$\times$10$^{-11}$ Torr.

\begin{figure*}[floatfix]
\begin{center}
\includegraphics[width=1.6\columnwidth,angle=0]{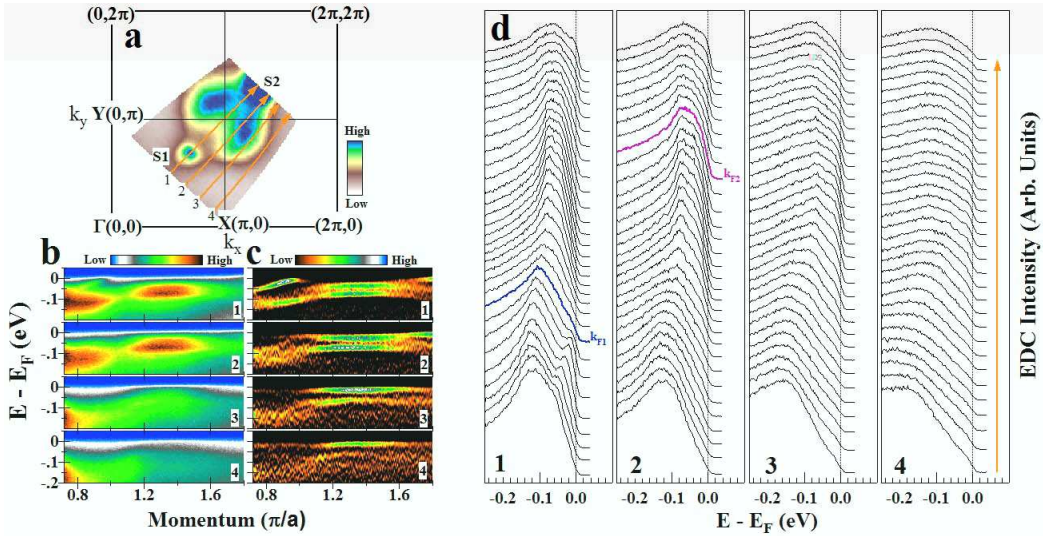}
\end{center}
\caption{Fermi surface, band structure and photoemission spectra of
(Sr,K)Fe$_2$As$_2$ near the M($\pi$,$\pi$) point. (a). Spectral
weight distribution integrated within [-5meV,5meV] energy window
with respect to the Fermi level as a function of k$_x$ and k$_y$.
(b). Original photoemission images measured along the four typical
cuts as marked in Fig. 2a. (c). Corresponding second derivative
images of Fig. 2b. (d). Photoemission spectra along the four cuts
with EDCs at the Fermi momenta colored and marked. }
\end{figure*}

Fig. 1 shows the Fermi surface (Fig. 1a), band structure (Figs. 1b
and 1c) and corresponding photoemission spectra (energy distribution
curves, EDCs)(Fig. 1d) on (Sr,K)Fe$_2$As$_2$ single crystal around
the $\Gamma$(0,0) point at a temperature of 45 K. The spectral
weight distribution integrated over a narrow energy window
[-5meV,5meV] near the Fermi level (Fig. 1a) gives a good
representation of the measured Fermi surface. Three Fermi surface
sheets can be clearly identified around the $\Gamma$ point from Fig.
1a, as marked in Fig. 3. The first is the well-defined inner small
Fermi surface sheet. The second is defined by nearly straight lines
that can be clearly seen in Fig. 1a. The third Fermi surface sheet
is defined by the outer strong intensity patches. We note that the
overall spectral weight distribution is not symmetrical with
four-fold symmetry with respect to the $\Gamma$ point. This is due
to photoemission matrix element effect because the main electric
field component of the light is along one particular diagonal
direction as marked in Fig. 1a.

Fig. 1b shows photoemission data taken along several typical
momentum cuts as marked in Fig. 1a. The corresponding second
derivative images, obtained by taking the second derivative on
photoemission spectra with respect to the energy at each momentum,
are shown in Fig. 1c. This is an empirical but effective way in
highlighting the underlying band structure\cite{WTZhang}. From Fig.
1c, three Fermi crossings can be seen, as marked by three arrows on
the data for the Cut 1, which correspond to three Fermi surface
sheets near the $\Gamma$ point. The measured bands further indicate
that all the three Fermi surface sheets are hole-like.  These band
dispersions and the Fermi crossings can also be seen from the
corresponding photoemission spectra  for the four cuts with the EDCs
at the Fermi crossings marked by colored lines (Fig. 1d). For the
inner small Fermi sheet, clear EDC peaks are observed at the
momentum crossings k$_{F3L}$ and k$_{F3R}$ in Fig. 1d for the Cuts 1
and 2.

Fig. 2 shows photoemission data of (Sr,K)Fe$_2$As$_2$ measured
around the M($\pi$,$\pi$) point at 45 K. The spectral weight
distribution (Fig. 2a) shows two strong intensity spots, S1 and S2,
along the $\Gamma$(0,0)-M($\pi$,$\pi$)-(2$\pi$,2$\pi$) line, with
their locations nearly symmetrical with respect to the
M($\pi$,$\pi$) point.  On both sides of the
$\Gamma$(0,0)-M($\pi$,$\pi$)-(2$\pi$,2$\pi$) line, there are two
patches of strong intensity. The maximum intensity contours on the
two patches are not enclosed and also the two strong intensity spots
appear to be isolated from the patches. These give rise to some
disconnected Fermi crossings identifiable around the M($\pi$,$\pi$)
point, as marked in Fig. 3. From the band structure measurements
(Fig. 2b and 2c), it is clear that the strong intensity spot, S1 in
Fig. 2a, originates from the band near the upper-left corner of
Figs. 2b and 2c for the Cut1.  From Fig. 2c, it is also clear that,
near the M($\pi$,$\pi$) point, the main electronic features are the
two flat bands which are $\sim$30 meV and $\sim$80 meV below the
Fermi level. The $\sim$30 meV band crosses the Fermi level and forms
an electron-like Fermi surface sheet near the M($\pi$,$\pi$) point
that correspond to the contour of the maximum intensity contour of
the two patches (as marked in Fig.3a near M point).

\begin{figure}[tbp]
\begin{center}
\includegraphics[width=1.0\columnwidth,angle=0]{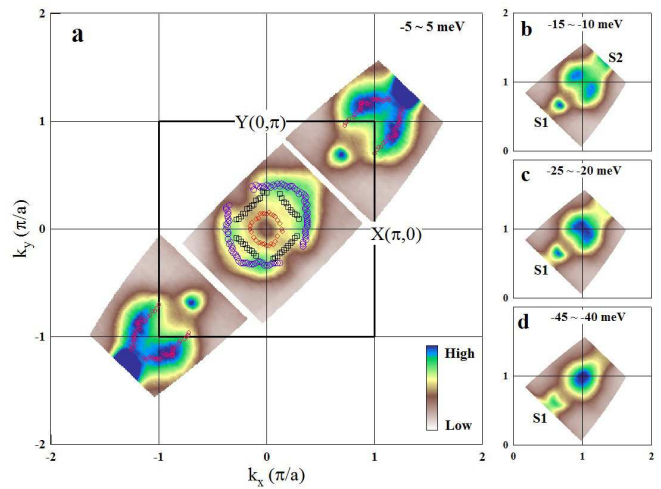}
\end{center}
\caption{Fermi surface of (Sr,K)Fe$_2$As$_2$. (a). Spectral weight
distribution integrated over a small energy window [-5meV,5meV] with
respect to the Fermi level.  The Fermi momenta are marked by
symbols. (b), (c) and (d) show spectral weight distribution near the
M($\pi$,$\pi$) point integrated over energy windows of
[-15meV,-10meV], [-25meV,-20meV] and [-45meV,-40meV], respectively,
with respect to the Fermi level.}
\end{figure}

Fig. 3 summarizes the overall Fermi surface of (Sr,K)Fe$_2$As$_2$ by
combining both measurements around the $\Gamma$ (Fig. 1a) and M
point (Fig. 2a).  Three hole-like Fermi surface sheets are resolved
around the $\Gamma$ point for the first time, different from two
Fermi surface sheets observed in (Ba,K)Fe$_2$As$_2$
compounds\cite{DLFKFe2As2,CLiuKFeAs}. Interestingly, the shape of
the three Fermi surface sheets appear to be not circular, but more
like squares. The enclosed area of the inner, middle and outer Fermi
sheets are $\sim$0.06, $\sim$0.28, $\sim$0.52, respectively, with a
unit of ($\pi$/a)$^2$. On the other hand, for the M point, there is
an electron-like Fermi surface sheet associated with the Fermi
crossings on the strong intensity patches (as marked near the M
point in Fig. 3a). However, because the patches are not enclosed
near the M point probably due to the matrix element effect, and the
appearance of two strong Fermi spots, S1 and S2, we need to further
determine whether the two strong spots and the patches are
independent or they belong to the same electron-like Fermi surface
sheet.  Fig. 3(b-d) shows spectral weight distribution integrated
over different energy ranges away from the Fermi level. The strong
Fermi spot S1 moves away from the M point with increasing binding
energy, which is consistent with the band dispersion as seen from
Fig. 2b and 2c (Cut 1, up-left band). On the other hand, the
electron-like Fermi surface defined by the patches gradually shrinks
towards the M point.  This clearly indicates that the two strong
Fermi spots S1 and S2 are independent from the electron-like Fermi
surface sheet defined by the patches near the M point.

Fig. 4 shows an overall band structure of (Sr,K)Fe$_2$As$_2$ along
typical high symmetry lines. This measurement, together with Fermi
surface information (Figs. 1, 2 and 3) makes it possible to have a
direct comparison with theoretical calculations. Since there is no
band calculations available on (Sr,K)Fe$_2$As$_2$ and it is expected
that the band structure of SrFe$_2$As$_2$ is similar to that of
BaFe$_2$As$_2$\cite{Nekrasov}, we take the band calculations of
BaFe$_2$As$_2$\cite{Nekrasov,CLiuKFeAs,TXiang} for comparison. The
observation of three hole-like Fermi surface sheets near the
$\Gamma$ point is consistent with the band
calculations\cite{Nekrasov}.  However, the overall measured band
structure and Fermi surface show significant difference from the
band calculated results\cite{Nekrasov,CLiuKFeAs,TXiang},
particularly around the M point.  As shown in Fig. 4c, four bands
are expected from band calculations near the M point within 0.6 eV
energy range, with two bands near -0.2 eV that give rise to two
electron-like Fermi surface sheets around the M
point\cite{Nekrasov}. This is quite different from the experimental
results where only two shallow flat bands are observed near
$\sim$-0.03 eV and $\sim$-0.08 eV and only one Fermi surface sheet
is identified. Moreover, the extra band along the $\Gamma$-M cut
(Fig. 4a) that gives rise to the strong Fermi spots in the Fermi
surface mapping (Fig. 1a) is not present in the band calculation
(Fig. 4c).

\begin{figure}[tbp]
\begin{center}
\includegraphics[width=1.0\columnwidth,angle=0]{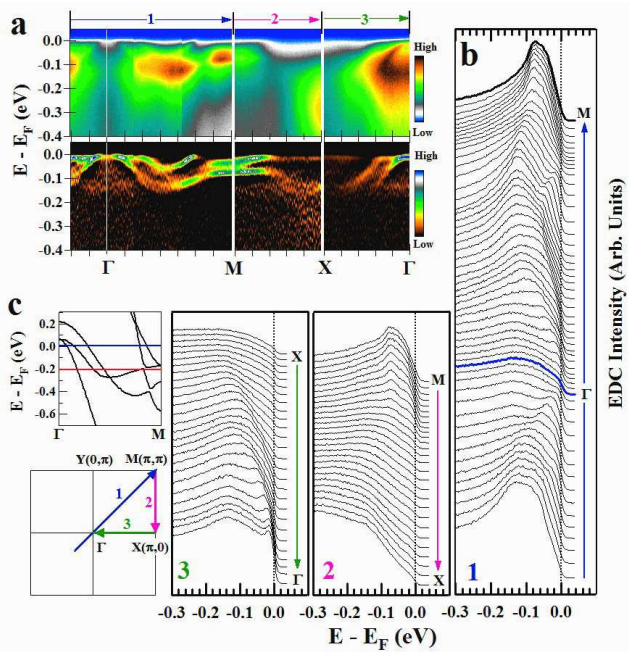}
\end{center}
\caption{Energy bands of (Sr,K)Fe$_2$As$_2$ along high-symmetry
lines. (a). Original photoemission images(upper panels) and the
corresponding second derivative images (low panels)\cite{HSNote}.
The locations of the momentum cuts are marked in the left-bottom
inset. (b). Corresponding photoemission spectra. The corresponding
momentum range is marked on top of Fig. 4a. (c). Calculated band
structure along the $\Gamma$-M direction in
BaFe$_2$As$_2$\cite{Nekrasov}. }
\end{figure}

The deviation between the measured and calculated electronic
structure may arise from a couple of reasons. First, there may be
appreciable uncertainty in the band calculation itself. For
instance, for the same BaFe$_2$As$_2$, the calculation by Liu et
al.\cite{CLiuKFeAs} gives only two Fermi crossings around the
$\Gamma$ point while three crossings are expected from calculations
of other groups\cite{Nekrasov,TXiang}.  Second, there can be some
uncertainty caused by k$_z$ dispersion. For a given photon energy as
we used, the measured electronic structure corresponds to one
particular k$_z$ with its location to be determined. However, we
believe this may not be the main reason because the bands near the M
point are not strongly sensitive to k$_z$\cite{Nekrasov,TXiang}.
Third, the effect of chemical potential shift has to be taken into
account. The potassium (K$^{+}$) doping in the (Sr,K)Fe$_2$As$_2$
sample introduces holes and is expected to lower the chemical
potential compared with the parent compound.  By carefully examining
the measured band structure (Fig. 4a) and the calculated one (Fig.
4c)\cite{Nekrasov}, we find that a qualitative agreement between the
measured and calculated electronic structure seems to be possible if
one assumes a chemical potential shift down by $\sim$0.2 eV (as
marked by a red line in Fig. 4c): (1). There are still three
hole-like Fermi surface sheets around the $\Gamma$ point; (2). There
is one extra band along $\Gamma$-M which resembles the one in Fig.
4a that gives rise to strong Fermi spots in Fermi surface mapping
(Fig. 2a); (3). Near the M point, this would lead to two bands with
one band crossing the Fermi level to give an electron-like Fermi
surface.

We note that while qualitative consistency may be realized by
shifting the chemical potential, there remain quantitative
discrepancies between the measured and calculated electronic
structures, such as the exact location of the Fermi crossings which
determines the area of the Fermi surface, that needs further
theoretical investigations.  One particularly notable difference
between the measurement and the calculations is the width of the
band dispersion. As seen from Fig. 4c, the major occupied bands
along $\Gamma$-M cut are spread within 0.6 eV energy range for
BaFe$_2$As$_2$, while they are within 0.15 eV in the measured data
(Fig. 4a). Even considering possible 0.2 eV chemical potential
shift, there remains obvious narrowing in the measured band width
compared with the calculated one. This appreciable band
renormalization suggests that electron correlation needs to be taken
into account in describing the electron structure of the Fe-based
compounds.

In summary, our angle-resolved photoemission measurements have
provided detailed electronic structure of the (Sr,K)Fe$_2$As$_2$
superconductor. Significant deviation between the measured and
calculated electronic structure is revealed that asks for further
theoretical efforts. The obvious bandwidth renormalization suggests
the importance of electron correlation in understanding the
electronic structure of the Fe-based compounds.

We thank Xiaogang Wen, Dunghai Lee, Zhong Fang, and Junren Shi for
helpful discussions. This work is supported by the NSFC, the MOST of
China (973 project No: 2006CB601002, 2006CB921302), and CAS
(Projects ITSNEM).

$^{*}$Corresponding author: XJZhou@aphy.iphy.ac.cn

\newpage


\begin{thebibliography}{99}
\bibitem{Kamihara}Y. Kamihara et al., J. Am. Chem. Soc. {\bf 130}, 3296
(2008).
\bibitem{XHChen43K} X. H. Chen et al.,  Nature {\bf 453}, 761 (2008).
\bibitem{GFChenCe} G. F. Chen et al., Phys. Rev. Lett. {\bf 100}, 247002 (2008).
\bibitem{ZARenNd}Z. A. Ren et al., Europhys. Lett. {\bf 82}, 57002 (2008).
\bibitem{ZARenPr}Z. A. Ren et al., arXiv:cod-mat/0803.4283.
\bibitem{ZARenSm}Z. A. Ren et al., Chin. Phys. Lett. {\bf 25}, 2215
(2008).
\bibitem{RotterSC}M. Rotter et al., arXiv:cod-mat/0805.4630.
\bibitem{Sasmal}K. Sasmal et al., arXiv:cod-mat/0806.1301.
\bibitem{GFCSrFeAs}G. F. Chen et al., arXiv:cod-mat/0806.1209.
\bibitem{XHCBaFeAs}G. Wu et al., arXiv:cod-mat/0806.1459.
\bibitem{Bednorz}J. G. Bednorz et al., Z. Phys. B {\bf 64}, 189 (1986).
\bibitem{PLeeReview}P. A. Lee et al., Rev. Mod. Phys. {\bf 78}, 17
(2006).
\bibitem{DongSDW}J. Dong et al., Europhys. Lett. {\bf 83}, 27006 (2008).
\bibitem{PCDai}C. Cruz et al., Nature {\bf 453}, 899(2008).
\bibitem{BFSSDW}M. Rotter et al., arXiv:cod-mat/0805.4021.
\bibitem{DJSingh} D. J. Singh and M.-H. Du, Phys. Rev. Lett. {\bf
100}, 237003 (2008).
\bibitem{HJZhang} H. J. Zhang et al., arXiv:cod-mat/0803.4487.
\bibitem{KHaule}K. Haule et al., Phys. Rev. Lett. {\bf 100},
226402(2008).
\bibitem{CCao}C. Cao et al., Phys. Rev. B {\bf 77}, 220506R (2008).
\bibitem{ZPYin}Z. P. Yin et al., arXiv:cond-mat/0804.3355.
\bibitem{FJMa}F. J. Ma et al., arXiv:cond-mat/0804.3370.
\bibitem{KaminskiLOFA}C. Liu et al., arXiv:cond-mat/0806.2147.
\bibitem{DLFKFe2As2}L. X. Yang et al., arXiv:cond-mat/0806.2627.
\bibitem{CLiuKFeAs}C. Liu et al., arXiv:cond-mat/0806.3453.
\bibitem{GDLiu} G. D Liu et al., Rev. Sci. Instruments {\bf 79}, 023105 (2008).
\bibitem{GFChenCrystal}G. F. Chen et al., arXiv:cod-mat/0806.2648.
\bibitem{Composition}The precise K content in the
(Sr$_{1-x}$K$_x$)Fe$_2$As$_2$ (T$_c$=21 K) is to be determined. From
the T$_c$$\sim$x curve\cite{Sasmal} it is estimated to be
x$\sim$0.25. It is underdoped compared with optimal doping at x=0.4
with a T$_c$=38 K.
\bibitem{WTZhang} W. T. Zhang et al., Phys. Rev. Lett. {\bf 101}, 017002 (2008).
\bibitem{HSNote} The X-M measurement is obtained from Y-M
meaurement which can be slightly different due to matrix element
effect from particular light polarization and sample measurement
geometry.
\bibitem{Nekrasov} I. A. Nekrasov et al., arXiv:cod-mat/0806.2630.
\bibitem{TXiang}F. J. Ma et al., cond-mat/0806.3526.

\end{thebibliography}
\end{document}